\documentclass[aps,prl,reprint,superscriptaddress,letterpaper]{revtex4-2}

\usepackage{amsmath}
\usepackage{amssymb}
\usepackage{braket}
\usepackage{mathtools}
\usepackage{bm}
\usepackage {verbatim}
\usepackage{upgreek}
\usepackage{xcolor}

\begin{document}

\title{Floquet engineering of strongly-driven excitons in monolayer tungsten disulfide}
\author{Yuki Kobayashi}
\thanks{These authors contributed equally}
\email{ykb@stanford.edu}
\affiliation{Stanford PULSE Institute, SLAC National Accelerator Laboratory, Menlo Park, CA 94025, USA}
\affiliation{Department of Applied Physics, Stanford University, Stanford, CA 94305, USA}

\author{Christian Heide}
\thanks{These authors contributed equally}
\affiliation{Stanford PULSE Institute, SLAC National Accelerator Laboratory, Menlo Park, CA 94025, USA}
\affiliation{Department of Applied Physics, Stanford University, Stanford, CA 94305, USA}

\author{Amalya C. Johnson}
\affiliation{Department of Materials Science and Engineering, Stanford University, Stanford, CA 94305, USA}

\author{Fang Liu}
\affiliation{Stanford PULSE Institute, SLAC National Accelerator Laboratory, Menlo Park, CA 94025, USA}
\affiliation{Department of Chemistry, Stanford University, Stanford, CA 94305, USA}

\author{David A. Reis}
\affiliation{Stanford PULSE Institute, SLAC National Accelerator Laboratory, Menlo Park, CA 94025, USA}
\affiliation{Department of Applied Physics, Stanford University, Stanford, CA 94305, USA}
\affiliation{Department of Photon Science, Stanford University, Stanford, CA 94305, USA}

\author{Tony F. Heinz}
\affiliation{Stanford PULSE Institute, SLAC National Accelerator Laboratory, Menlo Park, CA 94025, USA}
\affiliation{Department of Applied Physics, Stanford University, Stanford, CA 94305, USA}
\affiliation{Department of Photon Science, Stanford University, Stanford, CA 94305, USA}

\author{Shambhu Ghimire}
\email{shambhu@stanford.edu}
\affiliation{Stanford PULSE Institute, SLAC National Accelerator Laboratory, Menlo Park, CA 94025, USA}

\date{\today}

\maketitle

{\bf
Interactions of quantum materials with strong-laser fields can induce exotic nonequilibrium electronic states \cite{McIver20}.
Monolayer transition-metal dichalcogenides, a new class of direct-gap semiconductors with prominent quantum confinement \cite{Mak10}, offer exceptional opportunities toward Floquet engineering of quasiparticle electron-hole states, or excitons \cite{Wang18}.
Strong-field driving has a potential to achieve enhanced control of electronic band structure, thus a possibility to open a new realm of exciton light-matter interactions.
However, experimental implementation of strongly-driven excitons has so far remained out of reach.
Here, we use mid-infrared laser pulses below the optical bandgap to excite monolayer tungsten disulfide up to a field strength of 0.3 V/nm, and demonstrate strong-field light dressing of excitons in the excess of a hundred millielectronvolt.
Our high-sensitivity transient absorption spectroscopy further reveals formation of a virtual absorption feature below the $\bm{1s}$-exciton resonance, which is assigned to a light-dressed sideband from the dark $\bm{2p}$-exciton state.
Quantum-mechanical simulations substantiate the experimental results and enable us to retrieve real-space movies of the exciton dynamics.
This study advances our understanding of the exciton dynamics in the strong-field regime, and showcases the possibility of harnessing ultrafast, strong-field phenomena in device applications of two-dimensional materials.
}

%%%%%%%%%%%%%%%%%%%%%%%%%%%%%%%%%%%%%%%%%%%%%%%%%%%%%%%%%%%%%%%%
%Introduction
%%%%%%%%%%%%%%%%%%%%%%%%%%%%%%%%%%%%%%%%%%%%%%%%%%%%%%%%%%%%%%%%
%Intro 1, light control of TMD, strong-field control, unexplored but desirable
Laser-field manipulation of electronic band structure is at the frontier of quantum-materials research \cite{McIver20}.
In particular, monolayer transition-metal dichalcogenides (TMDs) exhibit remarkable excitonic properties as a result of their reduced dimensionality \cite{Mak10}, and recent spectroscopic studies have shown their promising potential toward novel optoelectronic devices \cite{Kim14,Sie15,Ye17,Yong18,Cunningham19,Yong19,Morrow20,Shi20,Earl20,LaMountain21}.
One of the common approaches is to coherently drive the interband transition by a near-resonant visible pulse and induce an AC-Stark shift to the exciton resonance \cite{Kim14,Sie15}.
Another possible yet largely unexplored approach is to drive excitons with a strong-laser field, thereby directly imparting momentum from the oscillating laser field to the electron-hole pairs.
This limit of light-matter interactions is known as the strong-field regime \cite{Brabec00}, in which unique laser-induced phenomena occur such as high-harmonic generation \cite{Ghimire11}, subcycle dynamic Franz-Keldysh effects \cite{Lucchini16}, and predicted control of topological properties \cite{Silva19}.

%Intro 2, critical field strength
To describe the strong-field regime of exciton dynamics, it is instructive to consider the critical-field strength ($F_\text{c}$) that is required to break apart an exciton into a free electron-hole pair.
In the dipole limit, the critical-field strength can be approximated as a ratio between the binding energy ($E_\text{b}$) and Bohr radius ($a_\text{B}$) such that $F_{\text{c}}=\frac{E_\text{b}}{e a_\text{B}}$.
Typical parameters of $E_\text{b}\approx300$ meV and $a_\text{B}\approx1$ nm \cite{Chernikov14} yield the critical-field strength of 0.3 V/nm (laser intensity of 12 GW/cm$^2$).
Previous studies have been limited below this order of the field strength; a recent study by Cunningham {\it et al.} reported the largest AC-Stark shift of 32 meV with a visible excitation, but the applied field strength was $<0.03$ V/nm ($<0.14$ GW/cm$^2$) \cite{Cunningham19}.

%Intro 3, experimental challenges
Driving and probing strong-field processes in two-dimensional materials poses several experimental challenges.
First, sample damage may occur as a result of real-carrier generation by irradiation of an intense laser.
Second, light-dressing effects are a coherent process that is induced only during the laser excitation, and probing cannot rely on incoherent optical processes such as photoluminescence.
Lastly, experimental signatures of the field-driven dynamics, particularly those from exciton states, are yet to be established.

\begin{figure*}[tb]
\includegraphics[scale=1.0]{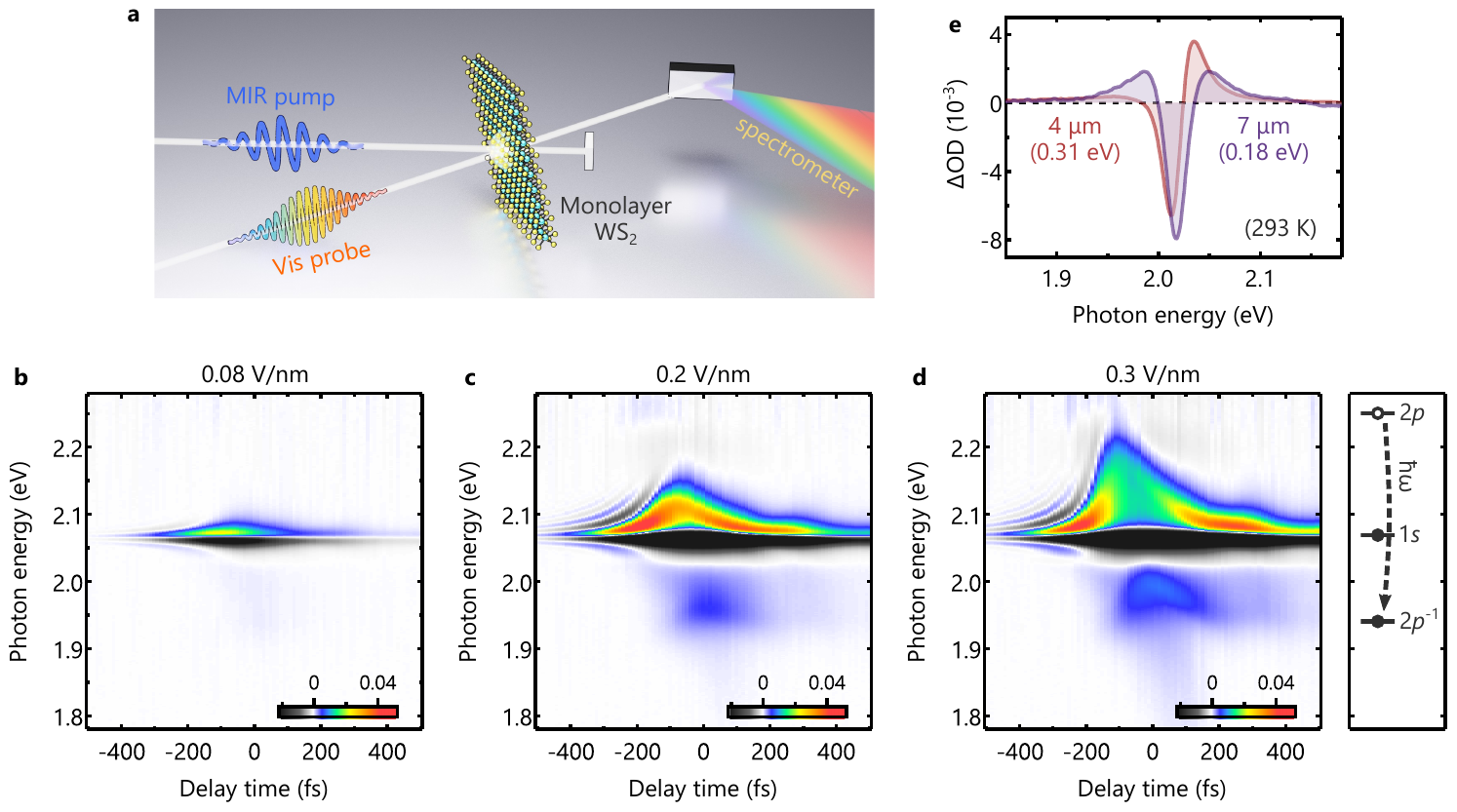}
\caption{\label{fig1}
{\bf Transient-absorption spectra of strongly-driven excitons in monolayer WS$_2$.}
{\bf a,} Outline of the experiments.
A mid-infrared pump pulse drives coherent exciton dynamics in a monolayer WS$_2$, and a visible probe pulse encodes the light-dressing effects in its transmission spectrum.
{\bf b-d,} Transient-absorption spectra ($\Delta$OD) of monolayer WS$_2$ at 77K driven by a 4-$\upmu$m mid-infrared field.
The positive (negative) delays correspond to the mid-infrared (visible) pulse arriving first.
Positive spectral chirp in the visible spectrum causes overall tilt of the transient-absorption signals, but this does not affect our main discussion.
At a field strength of 0.08 V/nm ({\bf b}), the light-dressing effect induces a blue shift of the $1s$-exciton signal via the AC-Stark shift.
As the intensity increases to 0.2 V/nm ({\bf c}), the AC-Stark shift grows, and a new absorption feature emerges below the $1s$-exciton resonance, which corresponds to a light-dressed sideband of the $2p$-exciton state, or the $2p^{-1}$-exciton state.
At the highest intensity of 0.3 V/nm ({\bf d}), significant blue shifts are characterized both for the $1s$- and $2p^{-1}$-exciton states.
The energy diagram on the right shows the estimated energies of the unperturbed light-dressed exciton states.
{\bf e,} Driving-wavelength dependence of the AC-Stark shift at the $1s$-exciton resonance at zero delay time.
The blue shift under the 4-$\upmu$m driving changes to a symmetric splitting under 7-$\upmu$m driving, which yields an estimated $1s$-$2p$ transition energy of 0.18 eV.
These measurements are performed at room temperature.
}
\end{figure*}

%Intro 4, what we did conceptually
Here, we overcome these challenges and demonstrate strong-field control of exciton states in monolayer tungsten disulfide (WS$_2$).
Our main experimental findings include an anomalously large energy shift to the $1s$-exciton resonance, and emergence of a light-dressed Floquet sideband from the $2p$-exciton state.
Experimentally, a 200-fs mid-infrared pulse of 4.0-um wavelength (photon energy of 0.31 eV, frequency of 75 THz) is used to excite monolayer WS$_2$ (Fig. 1a). 
The small photon energy of the mid-infrared pulse allows us to reach a field strength of 0.3 V/nm without damaging the sample, while light-dressing effects are induced by coherently driving internal resonance of the exciton states \cite{Poellmann15,Yong19}.
Such an intense mid-infrared pulse is produced by way of difference-frequency generation in a gallium-selenide crystal by using an output of a high-power optical-parametric amplification system.
We employ optical transient-absorption spectroscopy as a probing method, wherein a broadband visible pulse encodes the exciton dynamics in its transmission spectrum.
Differential absorption ($\Delta$OD) is obtained by taking the pump-on ($I_\text{on}$) and pump-off ($I_\text{off}$) spectra at a controlled delay time $t$, such that $\Delta\text{OD}=-\log(I_\text{on}/I_\text{off})$.
A high signal sensitivity is achieved by means of single-shot detection with a fast-rate spectrometer; accumulation of 30,000 laser shots leads to the detection sensitivity of $3\times10^{-5}$.
A high-quality, millimeter-scale sample of monolayer WS$_2$ is prepared by the gold-tape exfoliation method, and it was kept in a vacuum cryostat at 77 K \cite{Ajayi17,Liu20}.

%%%%%%%%%%%%%%%%%%%%%%%%%%%%%%%%%%%%%%%%%%%%%%%%%%%%%%%%%%%%%%%
%Experiment, delay scans
%%%%%%%%%%%%%%%%%%%%%%%%%%%%%%%%%%%%%%%%%%%%%%%%%%%%%%%%%%%%%%%%
Transient-absorption spectra of monolayer WS$_2$ under 4-$\upmu$m driving are shown in Figs. 1b-d.
At a lower mid-infrared intensity (Fig. 1b, 0.08 V/nm), the $1s$-exciton signal at 2.07 eV shows a blue shift.
This asymmetric behavior cannot be explained by peak broadening \cite{Shi20}, indicating that the underlying exciton light-matter interactions are a coherent process.
Additional measurements are carried out by changing the mid-infrared wavelength (Fig. 1e), and the transient-absorption signal is found to be symmetric under 7-$\upmu$m driving (photon energy of 0.18 eV).
The symmetric splitting (i.e., Autler-Townes splitting) signifies that the 7-$\upmu$m field is near-resonant to the intraexcitonic transition \cite{Yong19}, thus yielding an estimated $1s$-$2p$ transition energy of 0.18 eV.

At a higher intensity (Fig. 1c, 0.2 V/nm), a large ($\sim40$ meV) blue shift is observed at the $1s$-exciton resonance.
More importantly, a new absorption feature emerges below the $1s$-exciton resonance, which we assign to a single-photon-dressed Floquet sideband of the $2p$-exciton state, or the $2p^{-1}$-exciton state.
The original $2p$-exciton state is dipole forbidden from the ground state and does not directly contribute to absorption spectra \cite{Ye14}.
Single-photon dressing, however, can create a replica of the $2p$-exciton state that has the opposite parity.
Hybridization between the bright $1s$-exciton state can impart optical strength under a moderate driving intensity, thus creating a new virtual state that is present only during the mid-infrared driving.

There are two adjunct features that are worthy of note (Fig. 1c).
One is hyperbolic sidebands that appear around the $1s$-exciton signal in the negative delays (i.e., the visible pulse arriving first).
This is due to the perturbation of the free-induction decay at the $1s$-exciton resonance by the strong mid-infrared field.
Similar features are observed commonly in time-resolved spectroscopy \cite{Lindberg88}.
The other feature is possible implication of carrier generation by the strong mid-infrared field.
A long delay-time measurement shows that the differential absorption at the $1s$-exciton resonance at 3000 fs is as small as $0.8\times10^{-3}$ (Supplementary), and thus it is safe to assume that the effect of carrier generation is negligibly small.

As the driving intensity increases to 0.3 V/nm (Fig. 1d), the blue shift at the $1s$-exciton resonance starts to entail significant peak broadening, which is indicative of exciton ionization.
It is also characterized that the $2p^{-1}$-exciton signal experiences a large ($\sim40$ meV) blue shift by the intense mid-infrared field.
As we will explain, this is an unexpected result from the conventional two-level picture; if the AC-Stark shift causes a blue shift to the $1s$-exciton state, the paired $2p^{-1}$-exciton state should experience the same amount of energy shift to the opposite direction, thus a red shift.

%%%%%%%%%%%%%%%%%%%%%%%%%%%%%%%%%%%%%%%%%%%%%%%%%%%%%%%%%%%%%%%%
%Experiment, intensity scans and simulations
%%%%%%%%%%%%%%%%%%%%%%%%%%%%%%%%%%%%%%%%%%%%%%%%%%%%%%%%%%%%%%%%
\begin{figure}[tb]
\includegraphics[scale=1.0]{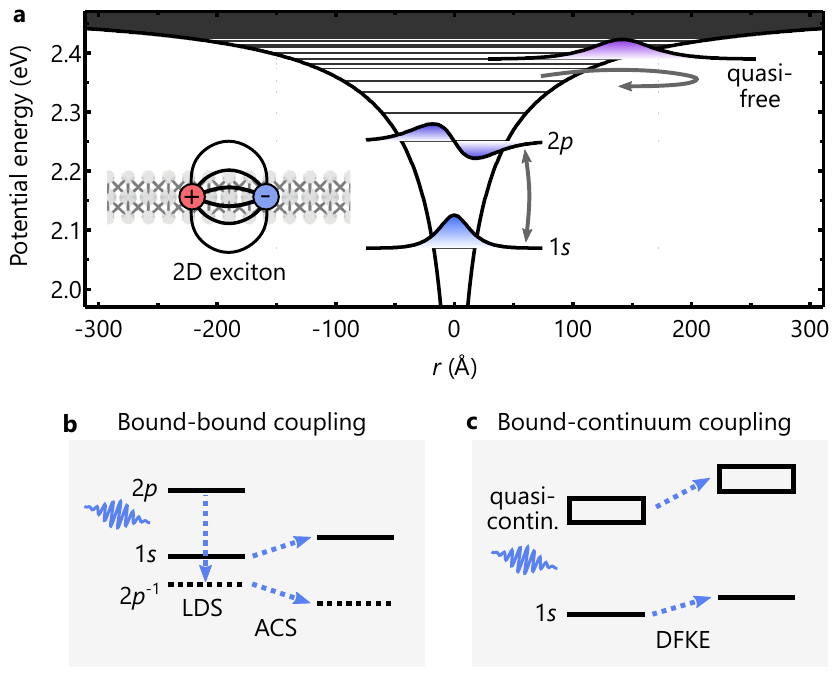}
\caption{\label{fig2}
{\bf Light-dressing mechanisms of strongly-driven excitons.}
{\bf a,} An effective two-dimensional potential of excitons and associated energy levels.
An external laser field can excite internal resonance of excitons such as the $1s$-$2p$ transition, or drive the exciton wavepacket into the quasifree region of the potential.
{\bf b,c,} Schematic illustration of the light-dressing effects.
In the bound-bound coupling case ({\bf b}), an external laser field can create light-dressed sidebands (LDS) and the mixing between the light-dressed states induces the AC-Stark shift (ACS).
In the bound-continuum coupling case ({\bf c}), excitons can gain kinetic energy from the external laser field in the quasifree region, and the renormalization of the energy levels results in the dynamic Franz-Keldysh effect (DFKE).
}
\end{figure}

We proceed to theoretical analysis of the experimental results.
Our primary interest lies in finding optical signatures of the field-driven dynamics of excitons.
The mechanisms of the mid-infrared light dressing are illustrated in Figs. 2a-c.
When the excitation is limited to the lower bound states, such as the $1s$- and $2p$-exciton states, the light-dressing effect can be understood within the Floquet theory (Fig. 2b) \cite{Shirley65,Sambe73}.
The external laser field creates replicas of the exciton states that are evenly spaced by the driving-photon energy, and hybridization between the neighboring states leads to the AC-Stark shifts \cite{Yong19}.
In the other case where a strong laser field is applied, exciton wavepackets can be driven out of the bound potential to the quasifree region (Fig. 2c).
The excitons in the asymptote can gain significant kinetic energy from the oscillating laser field, which leads to renormalization of energy levels (i.e., the dynamic Franz-Keldysh effect) \cite{Jauho96,Nordstrom98}. 
In the case of a parabolic band (i.e., quasifree electrons), the renormalization results in a universal blue shift of the energy structure by the amount equal to the ponderomotive energy \cite{Tsuji08}.

\begin{figure*}[tb]
\includegraphics[scale=1.0]{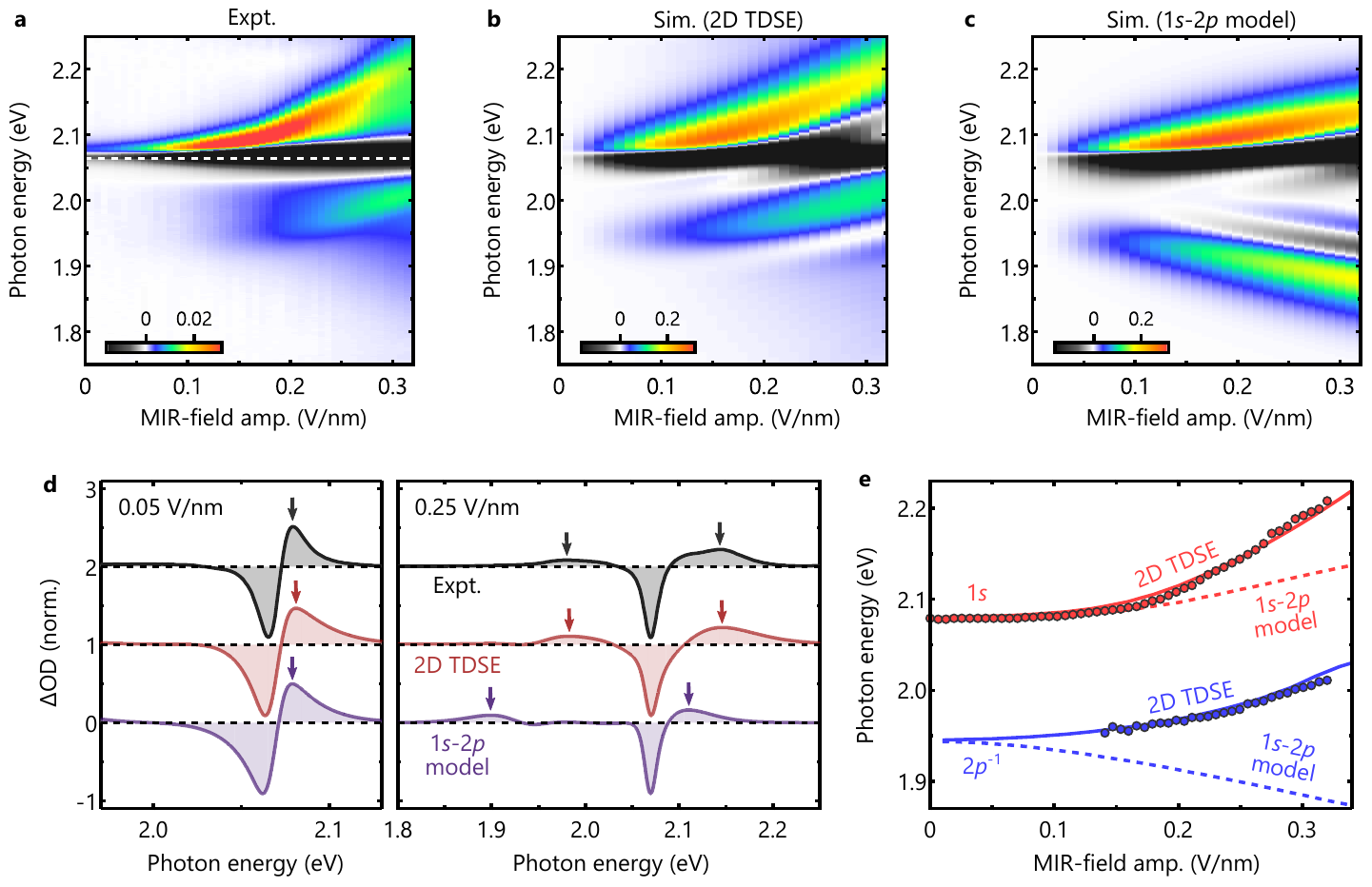}
\caption{\label{fig3}
{\bf Optical signatures of the field-driven exciton dynamics.} 
{\bf a-c,} Transient-absorption spectra ($\Delta$OD) of monolayer WS$_2$ driven by a 4-$\upmu$m mid-infrared field at various intensities.
The experimental transient-absorption spectra ({\bf a}) are measured at -60 fs and 60 fs delay times for above and below the $1s$-exciton resonance, respectively, to account for positive chirp in the visible spectrum.
The experimental results are compared to the simulation results obtained by the 2D-TDSE method ({\bf b}) and the $1s$-$2p$ model ({\bf c}).
{\bf d,} Comparison of the transient-absorption spectra at low (0.05 V/nm) and high (0.25 V/nm) mid-infrared field intensities.
The arrows indicate the peak energies of the $1s$- and $2p^{-1}$-exciton states.
{\bf e,} Summary of the experimental and simulated peak energies for the $1s$- and $2p^{-1}$-exciton states. 
}
\end{figure*}

We simulate transient-absorption spectra of the strongly-driven excitons with two models.
In the first model, the exciton light-matter interactions are fully considered, including the bound-bound coupling, bound-continuum coupling, and ionization.
This is achieved by numerically solving the time-dependent Schr{\"o}dinger equation (TDSE) for an exciton wavepacket in the nonhydrogenic two-dimensional Keldysh potential \cite{Keldysh79,Chernikov14},
\begin{align}
V(r)=-\frac{\pi e^2}{2 r_0}\left[H_0\left(\frac{r}{r_0}\right)-Y_0\left(\frac{r}{r_0}\right)\right],
\end{align}
where $H_0$ and $Y_0$ are Struve and Bessel functions, and $r_0$ is the screening length of Coulomb attraction in monolayer WS$_2$ (Fig. 2a).
The reduced mass of the exciton is fixed at $\mu=0.16m_0$ \cite{Berkelbach13}, and the free parameter of the screening length is chosen to be $r_0=50$ \AA{} so that the experimental $1s$-$2p$ transition energy of 0.18 eV is reproduced.
The other model uses a minimum discrete system that consists only of the ground, $1s$-, and $2p$-exciton states.
In this case, the mid-infrared field can drive only one bound-bound coupling, i.e., the $1s$-$2p$ transition, and the exciton dynamics in the quasifree region are excluded.
In both models, the absorption cross section of the exciton wavepacket is calculated as
$
\sigma(\omega)\propto\omega \operatorname{Im} \left[ \frac{\widetilde{d}(\omega)}{\widetilde{E}(\omega)} \right],
$
where $\widetilde{d}(\omega)$ and $\widetilde{E}(\omega)$ are the Fourier-transformed optical dipole moment and laser electric field, respectively (see Methods for more details) \cite{Wu16}.

Figure 3 shows the comparison of the experimental (Fig. 3a) and simulated (Figs. 3b,c) transient-absorption spectra at various mid-infrared intensities.
To account for the visible spectral chirp in the experiments, the transient-absorption signals below and above the $1s$-exciton resonance are measured at $60$ fs and $-60$ fs, respectively
In the low-intensity case at 0.05 V/nm (Fig. 3d), the AC-Stark shift is the main feature of the transient-absorption signals, and the two models reproduce the experimental result.
In the high-intensity case at 0.25 V/nm (Fig. 3d), the 2D-TDSE method captures the anomalously large blue shift in the 1$s$-exciton signal as well as the emergence of the light-dressed 2$p^{-1}$-exciton state. % at 1.98 eV.
In contrast, the $1s$-$2p$ model exhibits a poor matching; the energy of the $1s$- and $2p^{-1}$-exciton signals are underestimated by $\sim30$ meV and $\sim80$ meV, respectively.
Overall, the 2D-TDSE method successfully reproduces the $1s$- and $2p^{-1}$-exciton signals throughout the intensity regime measured in the experiments, whereas the $1s$-$2p$ model works only at the lower field strengths (Fig. 3e).
The contrast between the two models originates from the fact that the 2D-TDSE method fully considers the bound-bound and bound-continuum couplings that are expressed within the grid space ($r\leq300$ \AA{}), whereas the $1s$-$2p$ model is strictly restricted to the single bound-bound coupling.

%%%%%%%%%%%%%%%%%%%%%%%%%%%%%%%%%%%%%%%%%%%%%%%%%%%%%%%%%%%%%%%%
%AC-Stark Shift vs DFKE
%%%%%%%%%%%%%%%%%%%%%%%%%%%%%%%%%%%%%%%%%%%%%%%%%%%%%%%%%%%%%%%%

\begin{figure*}[tb]
\includegraphics[scale=1.0]{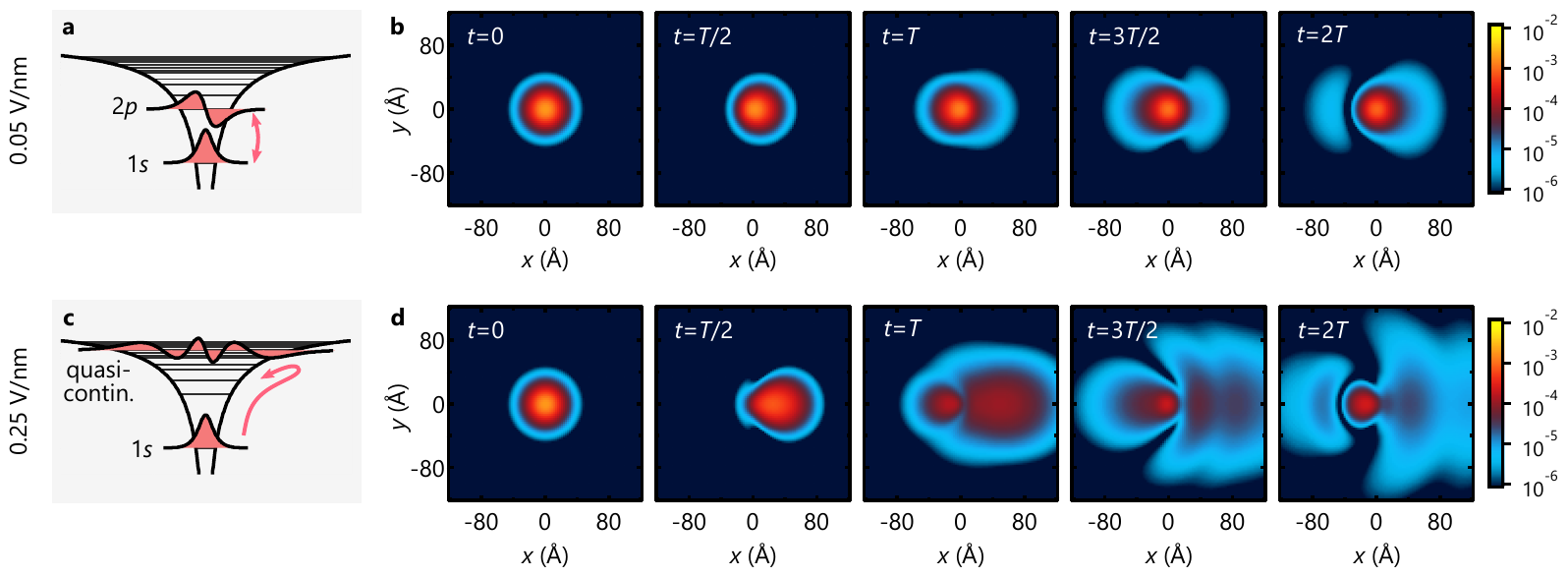}
\caption{\label{fig4}
{\bf Snapshots of the exciton wavepacket.} 
{\bf a} Schematic of the exciton dynamics in a low-intensity case (0.05 V/nm).
({\bf b}) Snapshots of the exciton wavepacket (squared probability distribution) obtained in the 2D-TDSE simulations.
The wavepacket remains localized at the potential center, and it exhibits only one node after two cycles ($t=2T$) of the mid-infrared excitation.
This shows that the excitation is limited to the $2p$-exciton state.
({\bf c}) Schematic of the exciton dynamics in a high-intensity case (0.25 V/nm).
{\bf d,} Here, the strong mid-infrared field drives the exciton wavepacket out of the bound potential into the quasifree region within a cycle.
The wavepacket is no longer localized at the potential center, and multiple nodes appearing in the snapshots represent the interference between the excited-state quasicontinuum.
}
\end{figure*}

The good agreement between the experiments and the 2D-TDSE simulation further allows us to obtain real-space pictures of the strongly-driven excitons (Fig. 4).
Here, the initial wavepacket is assumed to be the $1s$-exciton state launched by the visible pulse at $t=0$, and the mid-infrared driving field is a sine pulse (i.e., the electric field is zero at $t=0$).
In the low intensity case at 0.05 V/nm (Figs. 4a,b), the exciton wavepacket remains localized around the potential center.
After two cycles ($t=2T$), the exciton wavepacket exhibits only one node structure, which shows that the excitation is limited to the $2p$-exciton state.
In the high intensity case at 0.25 V/nm (Figs. 4c,d), the center-of-mass of the wavepacket is moved by $\sim30$ \AA{} only after a half cycle ($t=T/2$).
Multiple nodal features emerge at later times, which originate from interference between the excited-state quasicontinuum.
Note that a certain portion of the population remains at the potential center, which shows that we are in the regime where both the bound-bound coupling (i.e., AC-Stark shift) and the bound-continuum coupling (i.e., dynamic Franz-Keldysh effect) are in effect.
These observations, in combination with the results in Fig. 3, establish the role of the field-driven dynamics of excitons under the strong mid-infrared driving.

%%%%%%%%%%%%%%%%%%%%%%%%%%%%%%%%%%%%%%%%%%%%%%%%%%%%%%%%%%%%%%%%
%Outlook
%%%%%%%%%%%%%%%%%%%%%%%%%%%%%%%%%%%%%%%%%%%%%%%%%%%%%%%%%%%%%%%%
In conclusion, we investigated the strong-field regime of the exciton dynamics in monolayer tungusten disulfide by using an intense mid-infrared pump in a transient-absorption configuration.
We characterized an anomalously large ($\sim140$ meV) blue shift in the $1s$-exciton resonance and a completely new light-dressed sideband originating from the optically-dark $2p$-exciton state.
We reproduced our experimental results by solving the time-dependent Schr{\"o}dinger equation in a two-dimensional Keldysh potential, which subsequently allowed us to reconstruct the subcycle evolution of the exciton wavepackets.
At moderate intensities, the wavepacket oscillates with the field but remains localized at the potential center, whereas at the highest intensities, the wavepacket spreads significantly such that multiple nodal structures arise from interference between the bound wavepacket and the continuum.
Previously, subcycle control of electron motion was studied for atomic systems \cite{Chini13,Ott14}.
Now that the applicability of the same concept is established for two-dimensional materials, it might be possible to trigger and control active motion of excitons at the fundamental time scale of the driving-laser fields.
This will open the path toward atomically-thin electroabsorption modulators with operation frequencies in the petahertz regime \cite{Kruchinin18}.
Our results also suggest that strongly-driven excitons could affect the high-harmonic generation process in monolayer TMDs, where enhanced per-layer efficiency has been reported compared to the bulk \cite{Liu17}.

%%%%%%%%%%%%%%%%%%%%%%%%%%%%%%%%%%%%%%%%%%%%%%%%%%%%%%%%%%%%%%%%
\bibliographystyle{apsrev4-2.bst}
\bibliography{biblist}

%%%%%%%%%%%%%%%%%%%%%%%%%%%%%%%%%%%%%%%%%%%%%%%%%%%%%%%%%%%%%%%%
\section{Acknowledgments}
We thank Jiaojian Shi, Ignacio Franco, and Vishal Tiwari for discussion.
This work was primarily supported by the US Department of Energy, Office of Science, Basic Energy Sciences, Chemical Sciences, Geosciences, and Biosciences Division through the AMOS program.
F.L. was supported by the Terman Fellowship and startup funds from the Department of Chemistry at Stanford University.
Y.K. acknowledges support from the Urbanek-Chorodow Fellowship from Stanford University.
C.H acknowledges support from the W. M. Keck Foundation and the Humboldt Research Fellowship.

%%%%%%%%%%%%%%%%%%%%%%%%%%%%%%%%%%%%%%%%%%%%%%%%%%%%%%%%%%%%%%%%
\section{Author contributions}
Y.K. and C.H. performed the experiments.
A.C.J. and F.L. fabricated the sample.
Y.K. performed the simulations and analyzed the results.
F.L., D.A.R., T.F.H., and S.G. supervised the project.
All authors contributed to the preparation of the manuscript.

%%%%%%%%%%%%%%%%%%%%%%%%%%%%%%%%%%%%%%%%%%%%%%%%%%%%%%%%%%%%%%%%
\section{Competing interests}
The authors declare no competing interests.

%%%%%%%%%%%%%%%%%%%%%%%%%%%%%%%%%%%%%%%%%%%%%%%%%%%%%%%%%%%%%%%%
\section{Data availability}
The experimental and simulation data presented in this study are available from the corresponding author upon request.

%%%%%%%%%%%%%%%%%%%%%%%%%%%%%%%%%%%%%%%%%%%%%%%%%%%%%%%%%%%%%%%%
%%%%%%%%%%%%%%%%%%%%%%%%%%%%%%%%%%%%%%%%%%%%%%%%%%%%%%%%%%%%%%%%
%%%%%%%%%%%%%%%%%%%%%%%%%%%%%%%%%%%%%%%%%%%%%%%%%%%%%%%%%%%%%%%%
\clearpage
\section{Methods}
%%%%%%%%%%%%%%%%%%%%%%%%%%%%%%%%
\noindent {\bf Experimental details of optical transient-absorption spectroscopy.} 
The output from the Ti:sapphire amplifier (Evolution, Coherent Inc., 6 mJ, 45 fs, 790 nm, 1 kHz) is used to pump an optical-parametric amplification (OPA) system (TOPAS-HE, Light Conversion Inc.).
The signal ($\sim1300$ nm) and idler ($\sim1900$ nm) from the OPA are mixed in a GaSe crystal (Eksma Optics Inc., z-cut, 0.5 mm thick) for difference-frequency generation, and the generated mid-infrared pulse is spectrally filtered by a bandpass filter that is centered at 4.0 $\upmu$m (Thorlabs Inc., FB4000-500).
For the 7-$\upmu$m measurement, a different bandpass filter at 7.0 $\upmu$m (Thorlabs Inc., FB7000-500) is used and the incident angle to the GaSe crystal is adjusted for phase matching at 7 $\upmu$m.
A combination of wire-grid polarizers are used to clean the polarization of the mid-infrared pulse while enabling continuous tuning of the laser intensity.
The broadband visible pulse is produced by focusing the signal pulse to a 3-mm sapphire plate.
The mid-infrared pump and the visible probe are focused onto a sample by a 150-mm ZnSe lens and by a 100-mm off-axis parabola, respectively, in a noncollinear geometry with a crossing angle of $\sim25^\circ$.
The monolayer WS$_2$ is placed in an optical cryostat (ST-100, Lake Shore Cryotronics, Inc.) at a temperature of 77 K.
The transmitted visible spectra are collected by a CaF$_2$ lens and directed into a fast-rate spectrometer (Ocean Insight Inc., Ocean-FX).
The pump-on and pump-off spectra ($I_\text{on}$ and $I_\text{off}$) are recorded on a single-shot basis of 1 kHz, and the differential absorption $\Delta\text{OD}=-\log(I_\text{on}/I_\text{off})$ is obtained at controlled delay times.
The detection sensitivity of $3\times10^{-5}$ is achieved for accumulation of 30,000 spectra (i.e., 15,000 pump-on and pump-off spectra each).  
A monolayer WS$_2$ sample is exfoliated from bulk single crystal (HQ graphene) on a passivated fused silica substrate (100 um thick, double-side polished, MTI corporation) by the gold-tape exfoliation method \cite{Ajayi17,Liu20}.

%%%%%%%%%%%%%%%%%%%%%%%%%%%%%%%
\ \\
\noindent{\bf 2D-TDSE simulation.}
In the two-dimensional time-dependent Schr{\"o}dinger equation (2D-TDSE) method, the Hamiltonian of the exciton wavepacket is given by
\begin{align}
H(x,y,t)=-\frac{1}{2\mu}\frac{\partial^2}{\partial x^2}-\frac{1}{2\mu}\frac{\partial^2}{\partial y^2}-V(x,y)\nonumber\\
-\overrightarrow{E}(t) \cdot \overrightarrow{d}(x,y)-\frac{i \Gamma}{2},
\end{align}
where $\mu$ is the reduced mass of an exciton, $V(x,y)$ is the effective model potential, $\overrightarrow{E}(t)$ is the external laser field, $\overrightarrow{d}$ is the dipole moment, and $\Gamma$ is the phenomenological parameter to account for the finite lifetime of excitons.
Atomic units are used in this section.
As described in the main text, the reduced mass of the exciton is set to be $\mu=0.16 m_0$ as determined previously by density-functional-theory calculations \cite{Berkelbach13}.
The free parameter of the screening length is chosen to be $r_0=50$ \AA{} such that the experimental $1s$-$2p$ transition energy of 0.18 eV is reproduced.
The exciton binding energies reported for monolayer WS$_2$ range from 0.32 eV \cite{Chernikov14} to 0.7 eV \cite{Ye14}, and the value obtained with these parameters, 0.42 eV, falls within this range.
The phenomenological lifetime of the excitons is set to be $\Gamma=15$ meV, which corresponds to a $1/e$ lifetime of 44 fs, to match the experimental linewidth of the $1s$-exciton signal.
The mid-infrared laser is approximated as a 75-fs Gaussian pulse with a center wavelength of $4 \upmu$m, and the visible laser is approximated as a 1-fs Gaussian pulse with a center wavelength of $600$ nm.
In the calculations, the laser fields are assumed to be polarized along the $x$ direction.
The exciton wavepacket is represented by the sinc-function discrete variable representation (sinc-DVR) \cite{Colbert92} on $80\times80$ two-dimensional grids that span from -300 \AA{} to 300 \AA{}.
Complex absorption potential is defined at the grid boundary to prevent reflection of the wavepacket, which also models strong-field ionization of the excitons.
The time-dependent Schr{\"o}dinger equation is solved numerically by using the fourth-order Runge-Kutta method at a step size of 31 as up to 500 fs.
The absorption cross section of the exciton wavepacket is calculated as
\begin{align}
\sigma(\omega)\propto\omega \operatorname{Im} \left[ \frac{\widetilde{d}(\omega)}{\widetilde{E}(\omega)} \right],
\end{align}
where $\widetilde{d}(\omega)$ and $\widetilde{E}(\omega)$ are the Fourier-transformed optical dipole moment and laser electric field, respectively \cite{Wu16}.
The optical dipole moment $d(t)$ is considered proportional to the density of the wavepacket at $x,y=0$, which is a common approach used to calculate absorption spectra of semiconducting nanomaterials \cite{Glutsch96,Zhang06}.
A constant energy shift is added to the absorption cross section so that the position of the $1s$-exciton signal matches with the one measured in the experiments (2.07 eV).
For the simulation results presented in Fig. 3, the absorption spectra are averaged over two mid-infrared pulses, a cosine pulse and a sine pulse, to take into account the randomized carrier-envelope phase of the experimental mid-infrared pulses.

%%%%%%%%%%%%%%%%%%%%%%%%%%%%%%%%
\ \\
\noindent{\bf $\bm{1s}$-$\bm{2p}$ model simulation.}
In the $1s$-$2p$ model, the Hamiltonian of the exciton states is simplified as,
\begin{align}
H(t)=\left(
\begin{matrix} 0 & E(t) \cdot d_{g,1s} & 0 \\ E(t) \cdot d_{g,1s} & \epsilon_{1s}-i\Gamma/2 & E(t) \cdot d_{1s,2p} \\ 0 & E(t) \cdot d_{1s,2p} & \epsilon_{2p}-i\Gamma/2 \end{matrix}
\right),
\end{align}
where $\epsilon_i$ and $d_{i,j}$ represent the energy and transition dipole moment, respectively.
These state parameters are obtained from the effective model potential used in the 2D-TDSE method.
The same parameters and methods as in the 2D-TDSE method are used for the laser field, exciton lifetime, and time propagation.
The optical dipole moment is obtained as
\begin{align}
d(t)=\braket{\Psi(t)|\bm{d}|\Psi(t)},
\end{align}
where $\Psi_(t)$ is a column vector for the state coefficients and $\bm{d}$ is the transition dipole matrix.
The absorption cross section is calculated in the same way by using the Eq. (3).

%%%%%%%%%%%%%%%%%%%%%%%%%%%%%%%%%%%%%%%%%%%%%%%%%%%%%%%%%%%%%%%%
%%%%%%%%%%%%%%%%%%%%%%%%%%%%%%%%%%%%%%%%%%%%%%%%%%%%%%%%%%%%%%%%
\clearpage
\end{document}